# Enhanced Ionic Conductivity of confined Ionic-Liquid in Ångstrom-scale 2D channels

Jing Yang[1, 2, #], Raj Kumar Gogoi[3, 4, #], Chen Ming[5, #], Louis A. Maduro[3, 4, #], Abdulghani Ismail[3, 4], Hiran Jyothilal[3, 4], Kalluvadi Veetil Saurav[1, 4], Rongrong Qi[3, 4], Ravalika Sajja[3, 4], Ashok Keerthi[1, 4, 6], Robert A. W. Dryfe[1, 2], Alexei A Kornyshev[5, *], Boya Radha[3, 4, 6, *]

[1]Department of Chemistry, The University of Manchester, Manchester M13 9PL, U.K
[2]Henry Royce Institute, The University of Manchester, Manchester M13 9PL, U.K.
[3]Department of Physics and Astronomy, The University of Manchester, Manchester M13 9PL, U.K.
[4]National Graphene Institute, The University of Manchester, Manchester M13 9PL, U.K.
[5]Department of Chemistry, Faculty of Natural Sciences, Imperial College, Molecular Sciences Research Hub, White City Campus, London, W12 0BZ, UK
[6]Photon Science Institute, The University of Manchester, Manchester M13 9PL, U.K.

[#] Equally contributed
[*]Correspondence to: radha.boya@manchester.ac.uk and a.kornyshev@imperial.ac.uk

## Abstract

Understanding ion-transport under molecular confinement is essential for developing next-generation energy technologies, where ionic motion often occurs within nanoscale or Å-scale channels. In this study, we use the model system of 1-ethyl-3-methylimidazolium bis(trifluoromethanesulfonyl)imide ($[EMIM]^+[TFSI]^-$) confined within Å-scale slit-shaped 2D channels fabricated via van der Waals assembly (vdW) to exemplify a broader class of confined ionic liquids. This system provides a well-defined platform to unravel generic features of ion transport under extreme confinement. By systematically varying the channel height ($h$), we demonstrate a non-monotonic conductivity ($\sigma$) dependence on confinement, with $\sigma_{max}$ ~26.7 S.m$^{-1}$ at $h$ ~ 10.2 Å, over 30 times of the bulk value for these ionic liquids. The variation of $\sigma$ with confinement arises from structural rearrangements of ionic layers in the slit channel. Enhanced values of σ occur under confinements that promote the breakup of ion pairs and larger clusters, thereby increasing the number of 'free ions'. Stronger confinement ($h$ = 6.8 Å) also leads to steric hindrance, lowering $\sigma$ below bulk values. Furthermore, introducing co-solvents with a higher dielectric constant ($\varepsilon$) and lower viscosity ($\eta$), such as acetonitrile (ACN), amplifies $\sigma$ to ~145 S.m$^{-1}$. Comparative studies using ACN, dimethyl carbonate (DMC), and diethyl carbonate (DEC) highlight that both large $\varepsilon$ and low $\eta$ critically govern ion transport under confinement. The experimental results are supported by the molecular dynamics simulations by showing similar $h$-dependence qualitative trends, although they do not exactly reproduce the experimental $\sigma$-values. Overall, this work establishes confined $[EMIM]^+[TFSI]^-$ as a representative system for probing mechanisms of nano- and Å-scale ion transport, demonstrating how nanoconfinement and the solvent environment can be systematically tuned to manipulate ionic conductivity at the molecular level.

## 1. Introduction

Room-temperature ionic liquids (ILs), liquids composed entirely of cations and anions, have emerged as a distinctive class of electrolytes in both practical applications and fundamental research.[1] Their unique physicochemical properties such as non-volatility, high thermal and chemical stability, low flammability, and broad electrochemical windows, make them particularly attractive for energy-related technologies. ILs are considered promising candidates for use as electrolytes or additives in electrochemical double-layer capacitors (EDLCs) and have garnered significant attention in the broader field of energy storage. Although extensive research has studied the bulk properties of ILs including conductivity, electrochemical stability, transport properties, supramolecular structure, self-assembled solvent organization, and the free and bound states of ions;[1-3] their behaviour under nanoscale confinement has only recently emerged as a focus of study.[4-7] Over the past two decades, EDLCs based on nanoporous materials have attracted increasing interest due to remarkable improvements in energy density under strong confinement.[8-10] A growing research focus is on understanding how nanoscale confinement influences the transport properties of room-temperature ILs.[11] Although their bulk properties such as viscosity, diffusion, and ionic conductivity are well characterized,[1-3] the behaviour of ILs under nanoscale confinement in systems such as conical nanopores,[12,13] porous membranes,[14-16] and carbon-based matrices[17,18] reveal the effects of different pore geometry and surface chemistry of the pore walls. This opens an exciting and highly promising direction for both experimental and theoretical investigation, as understanding ionic motion within nanoscale or Å-scale channels is becoming essential for the development of next-generation iontronic technologies, including energy storage devices.

Recent advances in van der Waals (*vdW)* assembly techniques now enable the fabrication of Å-scale slit-shaped 2D channels with precisely tunable dimensions.[19,20] These platforms with Å-scale confinement present a unique opportunity to investigate ion transport in regimes dominated by molecular-scale packing, ion-wall interactions, strong Coulomb correlations, and steric effects. In such extreme confinements, classical continuum models often fail, giving rise to novel and unique transport phenomena such as ultra-fast water transport[19] and charge- and size-selective ion transport,[21,22] enabling ionic memristors,[23] molecular sensing,[24] and advanced spectroscopy.[25]

In this study, we systematically explore how Å-scale confinement and solvent environment affect the conductivity $\sigma$ of a room temperature IL, 1-ethyl-3-methylimidazolium bis(trifluoromethanesulfonyl) imide ($[EMIM]^+[TFSI]^-$), a well-studied one and known, in particular, for its excellent electrochemical stability. Using 2D channels with tunable $h$ ranging from sub-nm to 66 nm, we first examine $\sigma$ of the pure IL, revealing a non-monotonic dependence of $\sigma$ on confinement with a pronounced peak at ~10.2 Å, suggesting enhanced ion mobility under optimal confinement. Beyond confinement effects, the introduction of co-solvents offers an additional tool to modulate ionic transport. Solvents of high dielectric constant ($\varepsilon$) facilitate dissociation of ion pairs and higher order ion clusters, while low viscosity ($\eta$) ones enhance ion mobility. Complex interplay between confinement-induced structural ordering and solvent-mediated screening, largely unexplored earlier in concentrated ionic systems such as ILs, will be studied in this paper.

## 2. Materials and methods

**2.1. Fabrication of slit shaped 2D channels:** The slit-shaped 2D channels were fabricated by van der Waals (*vdW*) assembly of three layers of 2D materials, *viz*, top, bottom and spacers following our previously reported method.[19,20] Briefly, the bottom layer, the spacers, and the top layers are subsequently transferred onto a pre-made $SiN_X$ hole, followed by etching from the hole side to expose the channels, as shown in Fig. S1A. The top and bottom are 2D layers of hexagonal boron nitride (hBN), while the spacers were parallel strips (Fig. S1B, representative AFM image) of graphene layers. The number of layers is chosen depending on the requirements of the devices. A microscopic image of a trilayer stack is shown in Fig. S1C. A Cr(5nm)/Au(50nm) layer is deposited on top of the trilayer stack to enhance the adhesion of the channels with the $SiN_X$ substrate and to define the channel length ($l$). The height ($h$) of the 2D channels were determined by the number of graphene layers in the spacer flake (= 3.4 × number of graphene layers). The width ($w$) of the channels is defined by the space between the parallel array of spacers, generally 100 to 150 nm. The number of channels ($N$) for the devices are in the range of 200 ± 20.

**2.2. Synthesis of 1-ethyl-3-methylimidazolium bis(trifluoromethanesulfonyl) imide ($[EMIM]^+[TFSI]^-$):** The ($[EMIM]^+[TFSI]^-$) was synthesized following a procedure we previously reported.[26] Briefly, 75 g (0.5115 mol) 1-ethyl-3-methylimidazolium chloride and 154 g (0.54 mol) of lithium bis(trifluoromethanesulfonyl)imide were dissolved separately in 100 mL water. The two solutions were then mixed and stirred continuously at 40 °C overnight. The resulting mixture was transferred to a separating funnel and allowed to rest until two distinct phases formed. The bottom phase, consisting predominantly of the ionic liquid, was collected, while the top aqueous phase containing LiCl impurities was discarded. The collected ionic liquid phase was washed repeatedly with deionized (DI) water (at least twice the volume of the ionic liquid per wash). After thorough mixing, the mixture was transferred back to the separating funnel, and the ionic liquid phase was collected again. This washing process was repeated at least 10 times to ensure complete removal of LiCl impurities. Finally, the purified IL was dried under vacuum ($<6\times10^{-2}$ bar) at 70 °C for 3 days to remove residual water.

**2.3. Electrochemical cell setup and measurement protocol:** The electrochemical cell, constructed from polyether ether ketone (PEEK), features two small reservoirs designed to hold solutions and accommodate 2D channel devices. The devices are clamped between the reservoirs with the aid of two O-rings for sealing. Each reservoir has a capacity of ~0.4 mL and includes slots for electrode insertion. Prior to each measurement, the cell and O-rings were thoroughly cleaned using DI water and dried with a nitrogen gas flow. Once the 2D channel device was placed in the cell, either pure organic solvents or DI water, depending on the solution being tested, were injected into the cell to wash and wet both the cell and 2D channel devices. The voltage-dependent-current ($IV$) characteristics were measured using a Biologic® potentiostat (SP-300) equipped with an ultra-low current cable booster, offering a base current range from 1 µA to 1pA. The potentiostat low current resolution can reach 80 $a$A within the 1 pA range, as specified in the instructions. Data observed from the potentiostat were analysed using EC-lab software. Freshly prepared Ag/AgCl electrodes were used for each *I-V* measurement using two-electrode configuration where the reference and counter-electrode were connected to one Ag/AgCl electrode and the working electrode was connected to the other. The voltage scan rate was set at 40 mV $s^{-1}$ and when necessary, it was lowered to avoid capacitive effects that could mask the resulting current. All experiments were conducted at room temperature (293 K) inside a faraday cage.

**2.4. Channel height ($h$) dependent *I-V* measurements through slit-shaped 2D channels device.**
The effect of molecular confinement on the $\sigma$ of $[EMIM]^+[TFSI]^-$ was investigated by measuring the $IV$ characteristics in slit-shape 2D channels of different values of $h$. These 2D channels were fabricated by *vdW* assembly of 2D materials, primarily using hBN as the top and bottom walls, sandwiching layers of graphene strips as parallel spacers (schematic shown in Fig. S1).[19,20] The resulting trilayer stack was then placed over a pre-made micro-hole, forming the confined fluidic devices; schematic and microscopic images are in Fig. 1A and 1B, respectively. These devices offer excellent tuning of the channel dimensions even at the Å-scale level: length ($l$), width ($w$) and channels number ($N$) through nano-lithography process and $h$ determined by the selection of the number of graphene layers ($n$) in the spacers, following the relation $h = n \times 3.4$ Å. The *I-V* curves were recorded by sandwiching the device between two electrolyte reservoirs containing custom-made Ag/AgCl electrodes connected to a potentiostat (schematic represented in Fig. S5).

**2.5. Conductivity and conductance relationships:** The conductance of a 2D channel device is given by: $G = \sigma\left[\frac{whN}{l}\right]$ where $\sigma$ is conductivity, $w$ is channel width, $h$ is channel height, $l$ is channel length, and $N$ is the number of channels in the device.[21,27] The channel resistance is on the Gigaohm (GΩ) scale, which is much larger than the entry-exit resistance of the channel. Consequently, the conductance of ions is dominated by ion diffusion within the channel.

**2.6. Molecular dynamics (MD) simulations:** The MD simulations of the $[EMIM]^+[TFSI]^-$ inside the slit-shaped 2D channels are executed using the experimental setup represented in Fig. 2A. The channel size was defined as the slit height ($h$), following the experimental convention, and ranged from 3.3 Å to ~40 Å. The channel and reservoir lengths were set to 7 nm and 10 nm, respectively. Periodic boundary conditions were applied in all three directions. The channel walls (hBN surfaces in our case) were treated as rigid in all simulations, and B and N atoms were described using a Lennard-Jones potential[28], while the ionic liquid $[EMIM]^+[TFSI]^-$ was represented with an all-atom model.[29] The partial charges of cation and anion were scaled to 0.8 to account for the polarization. All simulations were carried out using the GROMACS software package.[30] In electrostatic interactions between charges representing ions and their interaction with hBN-walls, image forces due to dielectric polarizability of the hBN walls were not taken into account. The temperature of the IL was maintained at 300 K using the velocity-rescale thermostat within the canonical ensemble, where the number of particles (N), volume (V), and temperature (T) of the system are kept constant (NVT). Electrostatic interactions were computed using the particle mesh Ewald (PME) method[31,32], with a fast Fourier transform (FFT) grid spacing of 1 Å and a real-space cutoff of 1.2 nm. The systems were initially annealed from 400 K to 300 K over 5 ns, followed by an additional 15 ns of NVT simulation to achieve equilibrium. Subsequently, a further 10 ns of NVT simulation was performed for static property analysis. To improve statistical reliability, three independent simulations were conducted for each channel size.

The Green–Kubo (GK) equation has been used to estimate the $\sigma$ as it accounts for all interactions in dense ionic systems and provides a more reliable estimate than the Nernst–Einstein equation, which tends to overestimate the $\sigma$ of concentrated electrolytes.[3] The slit is confined along $z$ (~7 nm, *i.e.* length) but unbounded in $y$ under periodic boundary conditions, and the conductivity is anisotropic. Therefore, the reported $\sigma_{yy}$ reflects *in-plane* transport and allows direct comparison with experiments. Thus, from the GK- equation:

$$\sigma_{yy} = \frac{1}{Vk_bT}\int_0^\infty J_y(0)J_y(t)dt$$

where $V$ is the slit-volume, $k_b$ is the Boltzmann constant, and $T$ is the temperature, $J_y(t)$ denotes the component of the electric current along the $y$-direction at time $t$ (Fig. 2A). For each case, a 3 ns production simulation was performed, with trajectories saved every 0.05 ps for current autocorrelation function analysis. To improve statistical accuracy, fifteen independent simulations were carried out for each system, starting from different equilibrium configurations.

## 4. Results and discussions

Considering the ionic size of the $[EMIM]^+$ and $[TFSI]^-$ ion[33] (Fig. 1C), 2D channels with $h$ ranging from ~ 6.8 Å to 660 Å are used to investigate the confinement effect on the transport of the pure $[EMIM]^+[TFSI]^-$. The *IV*s through these 2D channels are linear, however slopes varied depending on $h$ (with other dimensions being constant), Fig. 1D. The bulk $\sigma$ of pure $[EMIM]^+[TFSI]^-$ varies depending on factors such as water content, temperature, and measurement methods, typically ranging between 0.5 and 1 $S.m^{-1}$.[34,35] We obtained a $\sigma$ ~ 0.87 $S.m^{-1}$ for pure liquid when measured using: (i) standard conductivity meter and (ii) measured through a micro-hole (25 μm × 2.5 μm × 500 nm). For the 2D channels with $h \geq 92$ Å, $\sigma$ is comparable to the bulk value. Increasing the confinement by reducing $h$ enhances $\sigma$ to a maximum of 26.7 $Sm^{-1}$ at $h$ = 10.2 Å (Fig. 1E). However, further confinement to $h$ = 6.8 Å results in $\sigma_{h\,=\,6.8\,Å} < \sigma_{bulk}$. In contrast, these 2D channels with aqueous electrolytes exhibit suppressed $\sigma$ values regardless of the type of ions used.[21,22,36] Thus, the confinement induced enhancement in $\sigma$ demonstrates the tunability of $\sigma$ of $[EMIM]^+[TFSI]^-$ with confinement, providing a route to manipulate ionic transport by precisely controlling $h$ at the atomic scale.

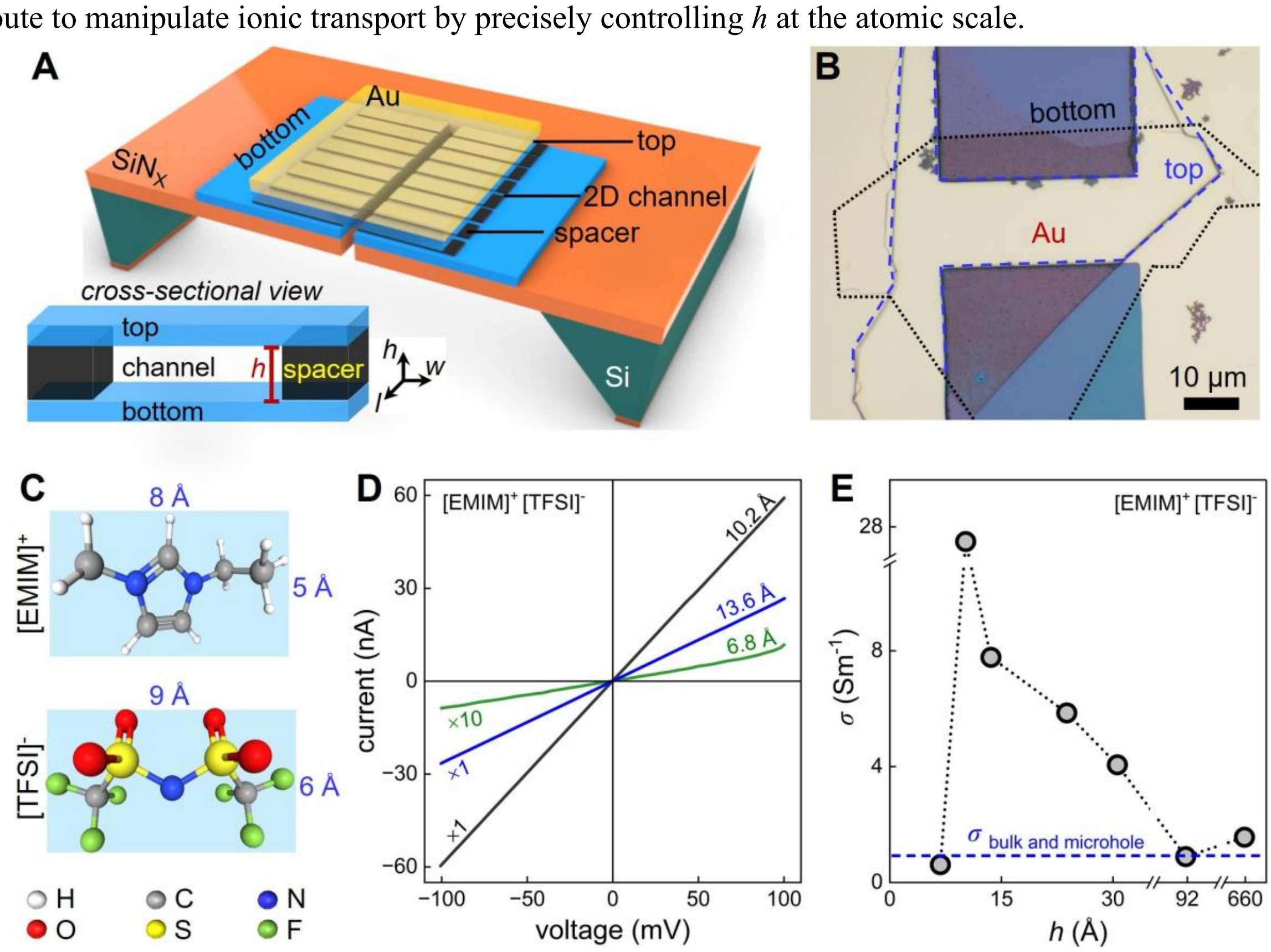

**Fig. 1: Confinement effect on $[EMIM]^+[TFSI]^-$ conductivity ($\sigma$).** (**A**) Schematic of the slit-shaped 2D channels on top of a micro-hole. Inset shows the cross-section view of a 2D channel. (**B**) Microscopic image showing the top-view of slit-shaped 2D channels device, where the spacers and 2D channels are sandwiched between the top and bottom layers. (**C**) Molecular structure and effective size of $[EMIM]^+$ and $[TFSI]^-$ ions. Comparison of (**D**) the voltage-dependent-current (*I-V*) curves of 2D channel devices with channel heights ($h$) of 6.8 Å, 10.2 Å and 13.6 Å. (normalized channels dimensions: length ($l$) = 1 µm, width ($w$) = 100 nm and number of channels ($N$) = 200) and (**E**) the $h$ dependent $\sigma$ of different channels, with the $\sigma_{bulk}$ and $\sigma_{microhole}$ using ionic liquid $[EMIM]^+[TFSI]^-$ as the electrolyte. In (**D**), the 6.8 Å plot is multiplied by 10 times for better comparison. The dotted line in (**E**) is a guide to the eye.

We continue with describing the MD simulations results of the $[EMIM]^+[TFSI]^-$ behaviour in the slit-shaped 2D channels to understand this confinement-enhanced $\sigma$ of $[EMIM]^+[TFSI]^-$. The calculated $\sigma$ (estimated by Green–Kubo relation, detailed in methods section) exhibits a non-monotonic dependence on $h$, with a maximum at $h$ ~ 10 Å (Fig. 2B), consistent with the experimental findings (Fig. 1E). Although the maximum occurs at the same height, it is less pronounced than in the experiments. Nevertheless, the similar qualitative trend suggests that the experimentally observed $\sigma$ enhancement originates from intrinsic ion dynamics under nanoscale confinement. To understand this behaviour, we first examined the *in-channel* ion structure by analysing number-density distributions (Fig. 2C), since the local organization of confined IL strongly influences their dynamics.[37] At $h$ ~ 3.3 Å, only a single mixed layer forms near the channel midplane. At $h$ ~10 Å, however, the ions reorganize into three distinct layers: cations preferentially located near the hBN walls due to stronger *vdW* interactions between the imidazolium ring and the surface, while an additional cation layer emerges at the centre, and anions populate a layer in between the two cation layers. As $h$ increases further, the central region becomes increasingly bulk-like. 2D number-density maps confirm that the emergence of this three-layer structure in the 10 Å-channel disrupts crystal-like ordering in the narrowest channel (Fig. 3), thereby enhancing ionic dynamics.

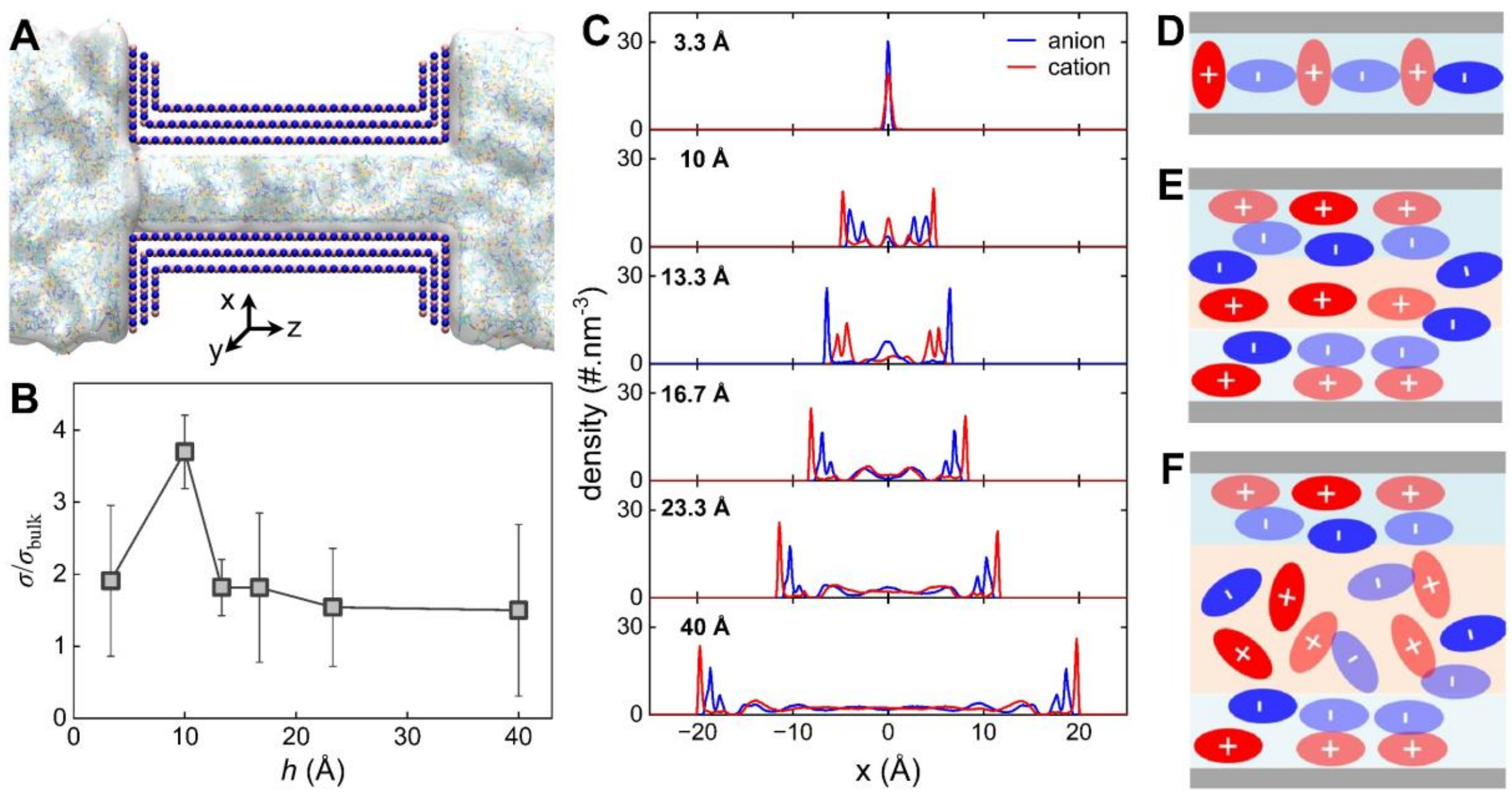

**Fig. 2: Molecular dynamics (MD) simulations.** (**A**) Schematics of MD simulation system setup. (**B**) $h$ dependent $\sigma$ enhancement of $[EMIM]^+[TFSI]^-$ estimated based on Green-Kubo equation, relative to the $\sigma_{bulk}$. (**C**) Ion number distribution along the $h$ of the channels. X = 0 represents the central plane between the channels' walls and the grey shaded areas, which represent the individual walls. Schematic of $h$ effect on the ion distribution; (**D**) 3.3 Å, (**E**) 10 Å, and (**F**) 23.3 Å.

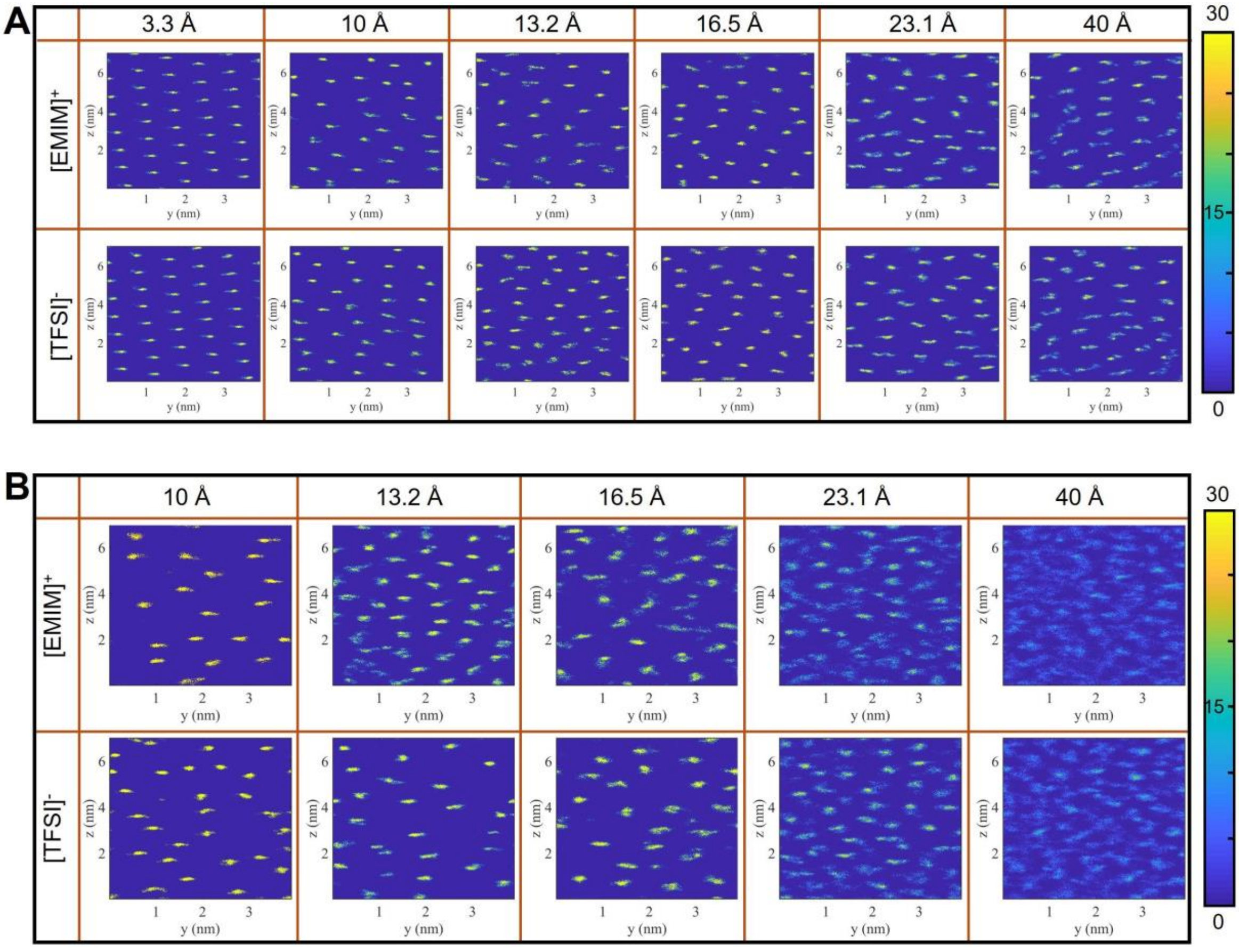

**Fig. 3: The 2D number density** of $[EMIM]^+$ and $[TFSI]^-$ inside the slit-shaped 2D channels (**A**) in the first interfacial layer, close to the wall and (**B**) in the middle region.

The confinement also affects the rotational degrees of freedom of the ions (Fig S7). The confinement elicits a channel-size dependent reorientation of both ions. In the narrowest slit ($h \sim 3.3$ Å), there is only one layer, where cations are predominantly perpendicular to the hBN plane and anions align in-plane. At intermediate $h$ (10 Å to 13.2 Å), the ring of interfacial cations reorient toward parallel to the plane, while anions develop a robust tilted subpopulation; The center at $h \sim 10$ Å mirrors this motif. Wider slits exhibit increasingly random, bulk-like orientations in the middle region, alongside a partial rebound of edge-on cations at the interface and a decline of tilted-anion fraction. Notably, the face-on (parallel) configuration at 10 Å is expected to weaken cation-anion interlocking and reduce electrostatic caging, thereby promoting ionic mobility and potentially enhancing the conductivity by enhanced *in-plane* motion.[10,37]

Further insight into the confinement-dependent conductivity is obtained from the free-ion population inside the channels. In IL, conductivity arises from the fraction of the free ions, while the majority of the cations and anions readily form ion pairs.[3] The in-channel number density of free ions increases from 2.08 $nm^{-3}$ at $h$ ~3.3 Å to ~2.30 $nm^{-3}$ at $h$ ~10 Å, then gradually decreases toward the stable value (Fig. S9A), displaying a nonmonotonic dependence on $h$, peaking at $h$ ~10 Å (19%). In addition, the diffusion coefficients (simulated for free ions *via* Green-Kubo relation) show an oscillatory dependence on $h$, reaching a maximum at $h$ ~10 Å for both cations and anions (Fig. S9B). Thus, $\sigma_{max}$ at $h$ ~10 Å originates from a cooperative structural and dynamical transition. In the narrowest slit, strong confinement enforces a single mixed layer with crystal-like ordering and the resulting steric hindrance severely limits ion transport and mobility. At $h$ ~10 Å, the system reorganizes into a three-layer (cation-anion-cation) arrangement, with ions adopting predominantly parallel orientations relative to the channel walls. This configuration weakens cation-anion interlocking, reduces electrostatic caging and steric constraints, and enhances free-ion diffusivity along the unconfined direction. Together with the peak in free-ion fraction, these effects account for the observed conductivity enhancement compared with other nano-channels.

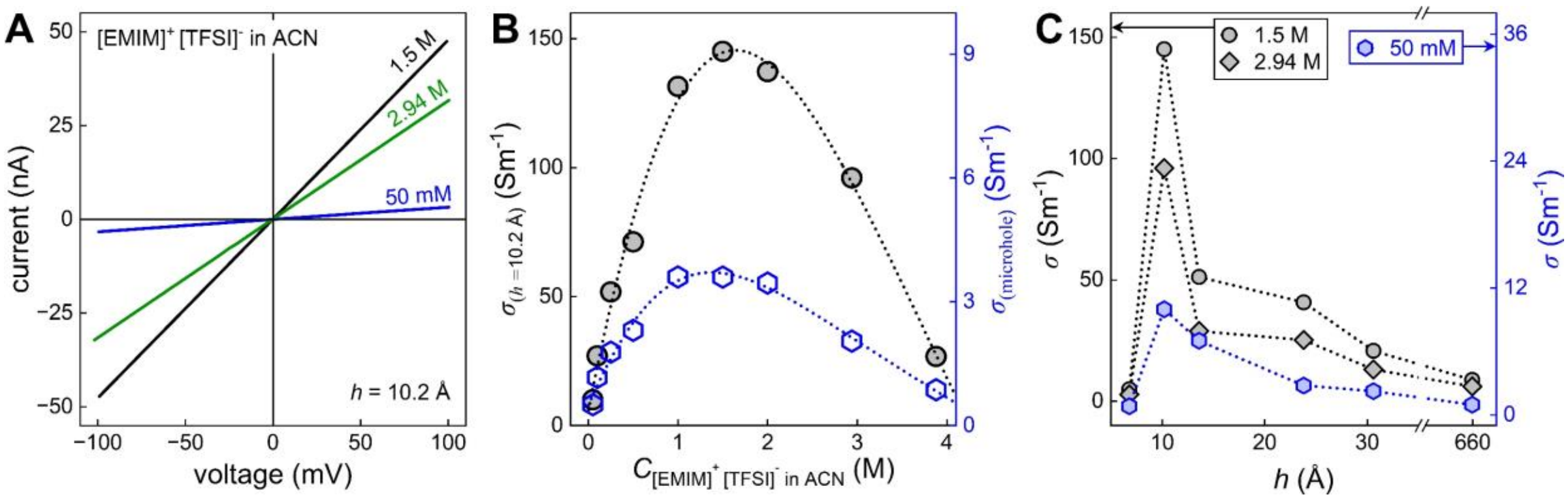

**Fig. 4: Effect of acetonitrile on $\sigma_{[EMIM]^+[TFSI]^-}$.** (**A**) Concentration $C$ dependent *I-V* curves of a $h$ = 10.2 Å slit-shape 2D channels device (channels dimensions: $l$ = 9.35 µm, $w$ = 145 nm and $N$ = 208). Comparison of the $\sigma_{[EMIM]^+[TFSI]^-}$ in ACN (**B**) of $h$ = 10.2 Å slit-shaped 2D channels and a micro-hole, and (**C**) of slit-shape 2D channels of various $h$ with $C_{[EMIM]^+[TFSI]^-}$ = 50 mM, 1.5 M and 2.94 M in ACN as electrolytes. The dotted lines in (**B**) and (**C**) are guides to the eye.

In IL-solvent mixtures, the $\sigma$ of an electrolyte depends on several factors including $\varepsilon$, $\eta$, $C$, diffusion coefficient and, critically the population of actual mobile charge carriers. The characteristics of the solvents used to prepare the electrolyte play a substantial role in determining its electrochemical properties. Solvents with higher $\varepsilon$ promotes more ion dissociation, increasing the number of free charge carriers and thereby increasing $\sigma$. The capacitance, $\sigma$, and $\eta$ of Ils in different organic solvents vary with dilutions.[38,39] To investigate the effect of confinement on $\sigma$ at different dilutions, we select ACN as a solvent, which has a higher $\varepsilon$ of 37.5, as compared to 12 of pure $[EMIM]^+[TFSI]^-$, and its strong dipole–dipole interactions, which promote preferential head-to-tail, antiparallel molecular arrangements and weak hydrogen bonding with [EMIM]$^-$.[39-43] For a 2D slit-channel device with $h$ = 10.2 Å, the addition of ACN initially enhances $\sigma$ increasing from ~26.7 S.m$^{-1}$ to a maximum of ~145 S.m$^{-1}$ upon diluting the IL from 3.88 M (*i.e.*, pure [EMIM]$^+$[TFSI]$^-$ concentration) to 1.5 M. Beyond this concentration, however, $\sigma$ decreases with further dilution, Fig 4B. This trend in $\sigma$ with dilution is also observed but with a different magnitude (~20 to 47 times, depending on the dilutions) (Fig 4B). Furthermore, we have investigated this dilution-dependent $\sigma$ of [EMIM]$^+$[TFSI]$^-$ under different degrees of confinement within the 2D channels, *viz*, $h$ (Å) = 6.8, 13.6, 23.8, 30.6 and 660. The dependence of σ as a function of $C$ exhibited similar behaviour for all confinement conditions, as shown in Fig. S11. However, the $\sigma$ measured at each $C$ followed the same trend as the confinement-dependent-$\sigma$ of pure [EMIM]$^+$[TFSI]$^-$; at each $C$, $\sigma$ increases as the confinement decreases, reaching a maximum at 10.2 Å and then decreases with further confinement, Fig 4C. This behaviour indicates the existence of an optimal confinement, arising from a balance between confinement-induced ordering and excessive steric restriction. To evaluate the influence of $\eta$ on $\sigma$, we then used dimethyl carbonate (DMC) as a solvent to dilute [EMIM]$^+$[TFSI]$^-$. DMC has a very low $\eta$ of 0.629 cP, significantly lower (~55 times) than that of pure [EMIM]$^+$[TFSI]$^-$ (~35 cP).[44] When [EMIM]$^+$[TFSI]$^-$ is diluted with DMC ($\varepsilon$ = 3.1), $\sigma$ initially increases from ~24 to 28.1 S.m$^{-1}$ for 2D channel device with $h$ = 10.2 Å as $C$ decreases from 3.88 to 2 M. Below 2 M, $\sigma$ begins to decrease with further dilution, Fig. 5A. A similar trend is observed in the micro-hole, though the overall $\sigma$ values are lower. This difference in $\sigma$ emphasizes the additional impact of geometric confinement on ion-transport of these electrolytic solutions, where slit channels provide more favourable conditions for directional and ordered ion migration compared to micro-hole slits. To further probe the role of solvent $\eta$ and $\varepsilon$ properties, we also studied [EMIM]$^+$[TFSI]$^-$ diluted in diethyl carbonate (DEC), which has a $\eta$ ~0.75 cP and ε ≈ 2.8. The $\sigma$ profile in DEC follows a similar trend as in DMC: down to $C$~2 M, $\sigma$ increases but then declines with further dilution (Fig. 5B). However, the maximum $\sigma$ values in DEC (Fig. 5B) are some 1.7 times smaller than those in DMC (Fig. 5A) under comparable confinement (partially due to slightly lower $\eta$ and higher ε; the difference in those values, however, is not that significant to entirely explain that effect).

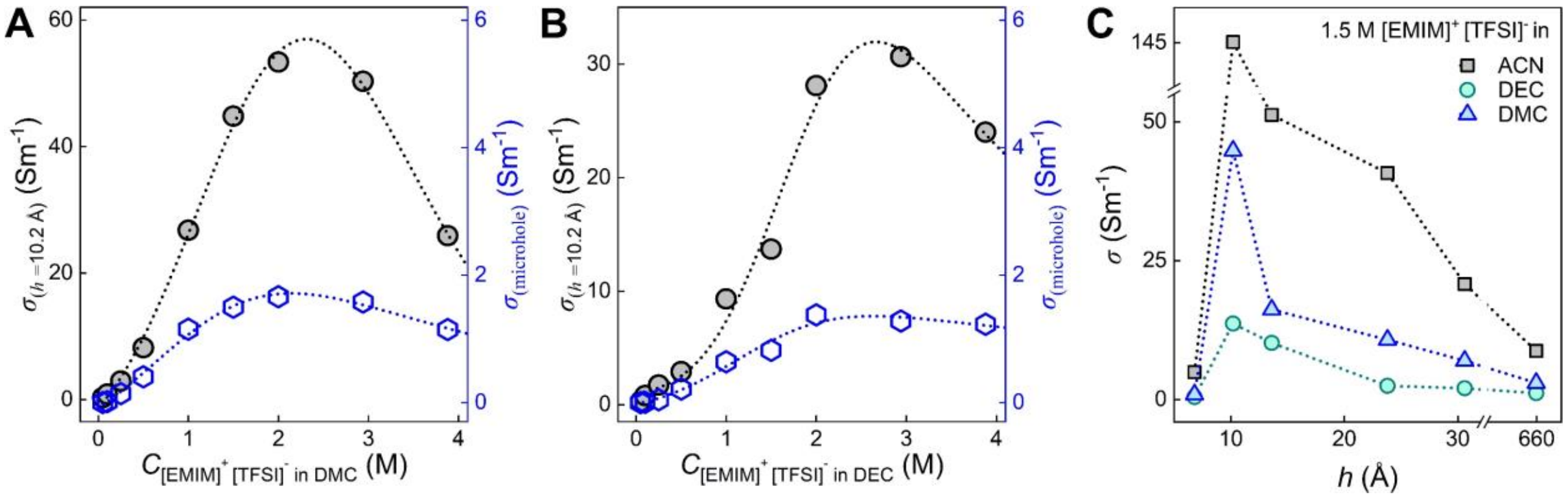

**Fig. 5: Effect of solvent on $\sigma_{[\text{EMIM}]^+[\text{TFSI}]^-}$.** Concentration dependent $\sigma_{[\text{EMIM}]^+[\text{TFSI}]^-}$ of $h$ = 10.2 Å slit-shaped 2D channels and a micro-hole with different solvents: (**A**) dimethyl carbonate (DMC) and (**B**) diethyl carbonate (DEC). (**C**) $h$-dependent $\sigma$ of slit-shaped 2D channels for 1.5 M [EMIM]$^+$[TFSI]$^-$ in ACN, DMC and DEC as solvents. The dotted lines in (**B**) and (**C**) are guides to the eye.

To understand the solvent-dependent σ of [EMIM]$^+$[TFSI]$^-$, MD simulations were performed with ACN, DMC, and DEC at various concentrations under bulk conditions. For the pure IL (~3.88 M), the fraction of free ions is low (~15.4%) due to strong Coulombic interactions promoting ion pairing.[3] Dilution increases the free ion fraction (Fig. S14A) as solvent molecules partially screen these interactions. ACN shows the most pronounced increase, reaching ~81.3% at 0.1 M, whereas DMC and DEC remain relatively low (~25% to 35%) across the concentration range, reflecting their limited dielectric screening. Consequently, ACN also exhibits the highest free ion density (Fig. S14B), DMC intermediate, and DEC the lowest, subject to their dielectric constants. Diffusion coefficients of both cations and anions increase with dilution as crowding and electrostatic correlations weaken. IL dissolved in ACN presents the highest mobility, followed by in DMC and DEC (Fig. S14C). The resulting ionic conductivity reflects the combined effects of free ion density and mobility: σ rises with initial dilution, peaks at intermediate concentrations, and then decreases at very low concentrations due to the limited number of charge carriers. ACN exhibits the highest $\sigma$ owing to its high dielectric constant and low viscosity, DMC shows intermediate values, and DEC the lowest, dominated by higher viscosity and weaker dielectric screening. These bulk trends provide a baseline for understanding confinement effects, as the same interplay of ion dissociation, mobility, and solvent properties govern $\sigma$ in 2D channels, where additional geometric restrictions can further enhance or modulate conductivity.

Fig. 5C presents the variation of σ with slit channel height at a fixed $[EMIM]^+[TFSI]^-$ $C$ of 1.5 M across the three solvents: ACN, DMC, and DEC. In all cases, $\sigma$ is highest at $h$ ~10.2 Å and gradually decreases as $h$ increases. At the smallest confinement of 6.8 Å, conductivity drops below the level observed in bulk conditions because the space becomes too tight for ions to move freely due to steric hindrance. This observation confirms that molecular confinement plays a critical role in determining ion transport. Among the solvents, $[EMIM]^+[TFSI]^-$ in ACN exhibits the highest $\sigma$ across all $h$, while DMC shows intermediate and DEC yields the lowest $\sigma$ values. This trend may arise because ACN combines a high $\varepsilon$ (37.5) with $\eta$ (0.334 cp), promoting efficient ion dissociation and high ionic mobility, whereas DMC has relatively low $\eta$ (0.629 cp) and moderate $\varepsilon$ (3.1), and DEC has a $\eta$ (0.75 cp) and lower $\varepsilon$ (2.8). However, further studies are required to understand the complex interplay of ionic liquids diluted in different solvents under confinement, since confining solvents in Å-scale channels can suppress the $\varepsilon$.[25] Nevertheless, ACN reinforces its role as an effective molecular 'lubricant,' easing ion motion in room temperature Ils– whether in the bulk or in nanoconfinement.

## 5. Conclusion

This study demonstrates how Å-scale confinement and organic solvent additives influence the $\sigma$ of $[EMIM]^+[TFSI]^-$. Confining the ionic liquid within 2D channels leads to a remarkable enhancement in $\sigma$, peaking at $h$ of ~10.2 Å due to confinement-induced mobility enhancement. However, when $h$ approaches the ionic diameter, steric hindrance reduces $\sigma$. Further modulation of $\sigma$ is achieved by introducing solvents such as ACN, DMC, and DEC, which affect ion dissociation and mobility through their $\varepsilon$ and $\eta$. Among them, ACN achieves the highest $\sigma$ values due to its favourable combination of high $\varepsilon$, which suppresses clustering of ions in IL, as well as low $\eta$. These findings highlight the effects of precise control over nanoscale geometry of the ion channels and electrolyte composition. By combining experiment and MD simulations, we established that the conductivity maximum originates from a confinement-driven transition in ion layering and orientation, which enhances free-ion fraction and in-plane diffusivity.

## Acknowledgements

B.R. acknowledges funding from the Royal Society University Research Fellowship URF\R\231008, Philip Leverhulme Prize PLP-2021-262, European Union's H2020 Framework Programme/ERC Starting Grant 852674 - AngstroCAP, EPSRC new horizons grant EP/X019225/1, Leverhulme Trust grant No. RPG-2025-136. A.K. acknowledges Royal Society grant ICA\R1\231014 and the Leverhulme Trust grant No. RPG-2026-190. B.R and A.K acknowledge EPSRC strategic equipment grant EP/W006502/1. R.A.W.D acknowledges EPSRC grant EP/V049925/1. A.A.K acknowledges the grant of EU Underwrite - Innovate UK EXANST PSP764 and the Leverhulme Trust grant No. RPG 2022-142. M.C and A.A.K acknowledge Prof. Guang Feng of Huazhong University of Science and Technology for useful discussions.

## Contributions

B. R. designed and directed the project; B.R., A.A.K, R.A.W.D and A.K. provided the overall supervision; J.Y. perfomed the ion transport measurements with help from A.I.; R.K.G., J.Y., A.I., analysed the ion transport measurements. L.A.M. led the 2D channels fabrication with help of H.J. A.K, K.V.S, R.Q and R.S.; C.M. and A.A.K. performed all the MD simulations; R.K.G and B.R. wrote the manuscript with J.Y., C.M. and A.A.K. All the authors contributed to the discussions

## Data availability

The data that support the findings of this study are available from the authors upon reasonable request.

## Conflict of interest

The authors declare no conflict of interest.